\documentclass[aps,prl,preprint,groupedaddress,showpacs,floatfix]{revtex4-1}

\usepackage{graphicx}
\begin{document}

\def\sn{{\rm sn}}
\def\cn{{\rm cn}}
\def\dn{{\rm dn}}
\def\am{{\rm am}}
\title{Isosynchronous Paths on a Rotating Surface}

\author{Neil Ashby}
\email[]{ashby@boulder.nist.gov}

\affiliation{University of Colorado,  Boulder, CO 80309-0390}
\altaffiliation{Present address:  Time And Frequency Division, Mail Stop 688, National Institute of Standards and Technology, Boulder, CO 80305}

\date{\today}

\begin{abstract}
In special relativity, clock networks may be self-consistently synchronized in an inertial frame by slowly transporting clocks, or by exchanging electromagnetic signals between network nodes.  However, clocks at rest in a rotating coordinate system--such as on the surface of the rotating earth--cannot be self-consistently synchronized by such processes, due to the Sagnac effect.  Discrepancies that arise are proportional to the area swept out by a vector from the rotation axis to the portable clock or electromagnetic pulse, projected onto a plane normal to the rotation axis.  This raises the question whether paths of minimal or extremal length can be found, for which the Sagnac discrepancies are zero.  This paper discusses the variational problem of finding such ``isosynchronous" paths on rotating discs and rotating spheres.  On a disc, the problem resembles the classical isoperimetric problem and the paths turn out to be circular arcs.  On a rotating sphere, however, between any two endpoints there are an infinite number of extremal paths, described by elliptic functions.
\end{abstract}

\maketitle

\section{Introduction}
		In an inertial reference frame, special relativity postulates that the speed of light has a universal value $c$ in all directions.  However in a rotating system of reference, such as on the rotating earth, this implies that light traversing a closed path will take different amounts of time to complete the circuit depending upon the path, and particularly upon whether the direction of propagation is in the general direction of, or opposite to, the rotation.  This effect is well known and is called the Sagnac effect.\cite{Post}

	It can be shown that when one slowly transports atomic clocks within a rotating system, an effect of the same form and magnitude occurs.  On earth, a portable clock transported slowly eastward once around the earth's equator will lag a master clock at rest on earth's surface by 207.4 nanoseconds (ns).  If transported westward the portable clock will lead by about 207.4 ns.  This effect was actually observed by Hafele and Keating,\cite{hafelekeating} who used commercial jet aircraft to transport an ensemble of Cesium clocks eastward and then westward around the globe.  

	For slowly moving portable clocks, the effect can be viewed from a local nonrotating geocentric reference frame, as being due to a difference between the time dilation of the portable clock and that of a master or reference clock whose motion is due only to earth's rotation.  For electromagnetic signals the effect can be considered to arise from a well-known consequence of special relativity--the relativity of simultaneity.  If one imagines two clocks fixed a small east-west distance $x$ apart on earth, then viewed from the nonrotating frame the clocks will be moving with approximately equal speeds $v = \omega r$, where $\omega$ is the angular rotation rate of earth and $r$ is the distance of the clocks from the rotation axis.  If a clock synchronization process were carried out by earth-fixed observers who ignored earth's rotation, then the two clocks would not be synchronous when viewed from the nonrotating frame.  The magnitude of the discrepancy is approximately $ v x/c^2 =(2 \omega/c^2)(rx/2)$.  Here, the factor $rx/2$ is the area swept out by a vector from the rotation axis to the portable clock or light pulse which accomplishes the synchronization, projected onto a plane normal to the rotation axis.\cite{ashby85}

	In general, such discrepancies depend on the path along which the synchronization process is conducted.  It is well known that the ``Sagnac discrepancy" $\Delta t$  can in general be expressed in terms of the integral\cite{landaulifshitz}
\begin{equation}
\Delta t = \left({2 \omega \over c^2} \right){1 \over 2}\int_{\rm path}(x\,dy - y\,dx),
\end{equation}
assuming the angular velocity vector is along the $z-$axis.  In terms of polar coordinates $(r,\phi)$ in the $(x,y)$ plane the Sagnac discrepancy is
\begin{equation}
\Delta t = \left({2 \omega \over c^2} \right){1 \over 2}\int_{\rm path}\, r^2 d\phi.
\end{equation}

	Out of the infinitude of paths from one point to another in a rotating coordinate system, there are trivial paths for which the projected area is zero.  For example, starting at some point on a rotating sphere one could slowly carry a portable clock along a meridian to the pole, then from the pole along another meridian to the final point.  Along each finite path segment, $d\phi = 0$, so the projected area vanishes.  A shorter path, still with vanishing projected area, can be obtained by detaching the path from the pole and allowing it to curve smoothly first in a direction opposed to the sense of rotation, then in the general sense of rotation, while passing near but not touching the pole.  This leads to the formulation of the question discussed in this paper:  What is the shape of a path of minimum length, for which the Sagnac discrepancy is zero?  Clearly such a path will not depend on the angular velocity's magnitude as the vanishing of the projected area will ensure that $\omega$ does not enter the solution.  We can therefore formulate the variational problem as one of finding the path of extremal length from an initial point $P_1$ to a final point $P_2$ in the rotating system, subject to the constraint that the area swept out by a vector from the axis of rotation to the path as it is described by the clock or electromagnetic signal, and projected onto a plane parallel to the equatorial plane, be zero.

	This problem resembles in some ways the classical isoperimetric problem of finding the shape of a closed curve of given length that encloses maximal area.  As is well known the solutions are circles and for the present problem the solutions turn out to be circular in shape but the vanishing of the projected area is enforced by combining areas both interior to and exterior to the circle.  For a path on a rotating sphere the solutions are expressible in terms of Jacobian elliptic functions, and have some unusual and interesting properties.  For want of a better term, we shall use the term ``isosynchronous" to describe these paths since vanishing of the projected area ensures that to the order of the calculation, clocks at the specified endpoints of the path will be synchronized by the slow transport of a portable clock or the transmission of an electromagnetic signal along the path.   

	We shall restrict our consideration here to two types of rotating systems: rotating discs and rotating spheres.  Section II formulates the variational problem for the disc in Cartesian and polar coordinates and develops a simple transcendental equation from which the radius of the solution circle may be determined.  We shall assume that the sense of a rotation from initial point $P_1$ to final point $P_2$ is in the general sense of rotation of the system.  

	Section III discusses the variational problem on a rotating sphere.  Study of this problem requires an elementary knowledge of the theory of Jacobian elliptic functions; the notation, as well as a summary of simple properties of these functions, is given in the Appendix.  As will be shown, the topology of the sphere permits infinite sets of extremal paths.  This is reminiscent of the infinite sets of paths of extremal length between two fixed points on the surface of a torus. 
	 
\section{Isosynchronous paths on a rotating disc}
\subsection{Rectangular Cartesian Coordinates}

	Using $(x,y)$ coordinates in the plane, the quantity to be extremized is 
\begin{equation}
W=\int_{P_1}^{P_2} \left(\sqrt{dx^2+dy^2}+{\lambda \over 2}(x dy - y dx)\right),
\end{equation}
where the length element along the path in the plane is
\begin{equation}
ds = \sqrt{dx^2+dy^2},
\end{equation}
$\lambda$ is a Lagrange multiplier and $P_1$ and $P_2$ are fixed initial and final points on the path of the synchronization process.  This expression can be written as a function of positions and ``velocities," by introducing a scalar parameter $t$ that varies along the path, and dividing and multiplying the integrand by $dt$.  Then
\begin{equation}
W=\int_{P_1}^{P_2} \left(\sqrt{\dot x^2+\dot y^2}+{\lambda \over 2}(x \dot y - y \dot x)\right)dt,
\end{equation} 
where a dot over a quantity indicates the derivative of that quantity with respect to $t$.  Thus effectively the Lagrangian for this case is 
\begin{equation}
\label{lagrangian1}
L=\sqrt{\dot x^2+\dot y^2}+{\lambda \over 2}(x \dot y - y \dot x).
\end{equation}
The Euler-Lagrange equations are then easily seen to be
\begin{eqnarray}
{d \over dt}\left({dx \over ds} - {\lambda \over 2}y \right) - {1 \over 2} \lambda \dot y=0,
\\
{d \over dt}\left({dy \over ds} + {\lambda \over 2}x \right) + {1 \over 2} \lambda \dot x=0.
\end{eqnarray}
These equations of motion can be integrated once yielding
\begin{eqnarray}
\label{circeqs1}
{dx \over ds}-\lambda(y-y_c)=0,
\\
\label{circeqs2}
{dy \over ds}+\lambda(x-x_c)=0,
\end{eqnarray}
where $(x_c,y_c)$ are integration constants.  Eqs. (\ref{circeqs1},\ref{circeqs2}) are equations of a circle with center at $(x_c,y_c)$.  One form for the solutions is:
\begin{equation}
\label{solutions1}
x=x_c+R \cos\left(\lambda(s-s_0)\right),\quad\quad y=y_c-R \sin\left(\lambda(s-s_0)\right).
\end{equation} 
$R$ and $s_0$ are integration constants.  $R$ is the circle's radius and the quantity $s_0$ represents the initial value of the path length $s$ along the circle.
\subsubsection{Boundary conditions}

		When the initial point $P_1=(x_1,y_1)$ and final point $P_2=(x_2,y_2)$ on the path are specified, the integration constants can be determined.  The unknowns are $x_c,\ y_c,\ R,\ \lambda(s_1-s_0),$ and $\lambda(s_2-s_0)$.  Also, the Sagnac effect must vanish, which is equivalent to vanishing of the area swept out along the path:
\begin{equation}
\label{zeroarea}
\int_{s_1}^{s_2}\left(x{dy \over ds}-y{dx \over ds} \right)\,ds =0. 
\end{equation}
Using the solutions given in Eqs. (\ref{solutions1}), this condition can be written
\begin{eqnarray}
\label{zeroarea2}
-R^2 \lambda(s_2-s_1) - R\bigg(x_c\left(\sin\left[\lambda(s_2-s_0)\right]-\sin\left[\lambda(s_1-s_0)\right]\right)\nonumber
\\ + y_c\left(\cos\left[\lambda(s_2-s_0)\right]-\cos\left[\lambda(s_1-s_0)\right]\right)\bigg)=0.
\end{eqnarray} 
A more compact form results if one introduces the angle $\theta=\lambda(s-s_0)$; in fact it will be shown that $\lambda=1/R$, so that $\theta$ measures the angle of an arc on the circle subtended by the center of the circle.  Thus if $\theta_2=\lambda(s_2-s_0)$ and $\theta_1=\lambda(s_1-s_0)$, then Eq. (\ref{zeroarea2}) becomes
\begin{equation}
-R^2(\theta_2-\theta_1)-R\bigg(x_c\left(\sin \theta_2 - \sin \theta_1\right)+y_c\left(\cos \theta_2 -\cos \theta_1\right)\bigg) = 0.
\end{equation}
The parametric solutions, Eqs. (\ref{solutions1}), give for the endpoints the conditions
\begin{eqnarray}
x_1=x_c+R \cos \theta_1;\nonumber\\
y_1=y_c-R \sin \theta_1;\nonumber\\
x_2=x_c+R \cos \theta_2;\\
y_2=y_c-R \sin \theta_2\,,\nonumber
\end{eqnarray}
and these can be combined to yield
\begin{equation}
\label{trianglearea}
x_1y_2-x_2y_1=-R^2 \sin(\theta_2-\theta_1)-R\left(x_c(\sin \theta_2-\sin \theta_1)+y_c(\cos \theta_2 -\cos \theta_1)\right)
\end{equation}
The linear term in radius $R$ in Eq. (\ref{zeroarea2}) can now be eliminated by using Eq. (\ref{trianglearea}):
\begin{equation}
\label{trianglearea1}
x_1 y_2-x_2 y_1= -R^2 \sin(\theta_2-\theta_1)-R^2(\theta_2-\theta_1).
\end{equation}
\begin{figure}[ht]
\includegraphics[width=3.125truein]{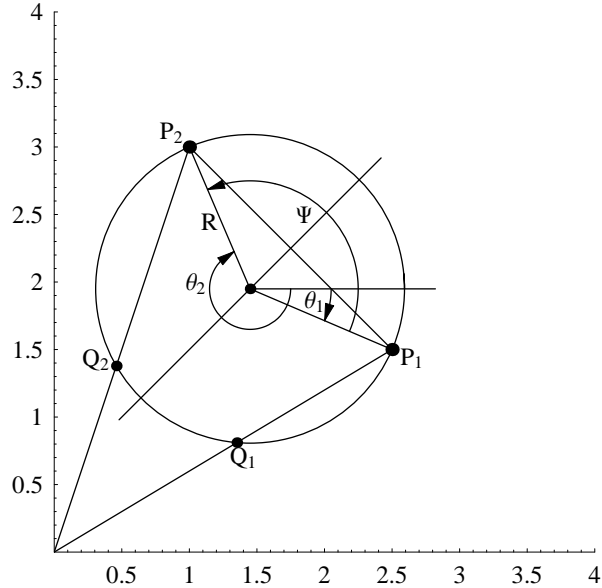}
\caption{Solution for $P_1=(2.5,1.5),\  P_2=(1,3)$.  The angles $\phi_1$ and $\phi_2$ are measured positive clockwise.  The points $Q_i$ are the intersections of the solution circle with straight lines from the rotation axis to $P_i$, respectively.  Here $\Delta=\theta_2-\theta_1=3.901 {\rm\ radians}, R=1.1427$.} 
\end{figure}
	Referring now to Fig. 1, the center of the circular path lies on the perpendicular bisector of the straight line from $P_1$ to $P_2$.  When $s=s_0$, $y=y_c$ and $x=x_c+R$ so the reference point for the measurement of angles lies on the circle, directly to the right of the center.  Let the distance between the specified endpoints be
\begin{equation}
d = \sqrt{(x_2-x_1)^2+(y_2-y_1)^2},
\end{equation}
and let $\Psi=2\pi -(\theta_2-\theta_1)$ be the angle subtended by the chord $P_1P_2$ at the center of the circle. Then 
\begin{equation}
{d \over 2} = R \sin(\Psi/2),
\end{equation}
so 
\begin{equation}
\label{Rbyd}
R^2 = {d^2 \over 4 \sin^2(\Psi/2)} = {d^2 \over 2\left(1-\cos(\theta_2-\theta_1)\right)}.
\end{equation}
Eliminating $R$ from Eq. (\ref{trianglearea1}) and rearranging gives
\begin{equation}
 {2(x_1y_2-x_2y_1) \over d^2} = {\Delta -\sin\Delta \over 1 - \cos\Delta}.
\end{equation}
where $\Delta = \theta_2-\theta_1$.  The left side of this transcendental equation involves only coordinates of the endpoints; the right side involves only the angle difference $\Delta$.  Given the endpoints, the equation is easily solved numerically for $\Delta$ by a method such as Newton's method, which converges rapidly since the function is monotonic.  Once $\Delta$ is found, the radius $R$ is found from Eq. (\ref{Rbyd}).

Let $l$ be the distance from the midpoint of the chord $P_1P_2$ to the center of the circle.  Then
\begin{equation}
l = R \cos (\Psi /2)=-R \cos(\Delta/2)= {d \over 2 \tan(\Psi/2)}.
\end{equation}
The center of the circle can now be located along the chord's normal through the midpoint of the chord.  The slope of the normal is $-(x_2-x_1)/(y_2-y_1)$.  Let the angle this normal makes with the horizontal axis be denoted by $\Phi$.  Then
\begin{equation}
\cos\Phi = {y_2-y_1\over d};\quad\quad \sin\Phi = -{x_2-x_1 \over d}.
\end{equation}
Then the $x-$coordinate of the center is
\begin{equation}
x_c = {x_1+x_2 \over 2} - l \cos\Phi = {x_1+x_2 \over 2}-{y_2-y_1 \over 2 \tan(\Psi/2)}. 
\end{equation}  
Similarly for the $y-$coordinate of the center,
\begin{equation}
y_c = {y_1+y_2 \over 2} - l \sin\Phi = {y_1+y_2 \over 2}+{x_2-x_1 \over  2 \tan(\Psi/2)}.
\end{equation}
The angles $\theta_1$ and $\theta_2$ are finally found from the boundary conditions;
\begin{eqnarray}
{x_1-x_c \over R } = \cos \theta_1; \quad\quad {y_1-y_c \over R} = -\sin\theta_1;\\
{x_2-x_c \over R } = \cos \theta_2; \quad\quad {y_2-y_c \over R} = -\sin\theta_2.
\end{eqnarray}

	Lastly, the Lagrange multiplier is determined by observing that $s$ is a length along the circle.  Clearly in order that $\theta_1$ and $\theta_2$ be angles in radians as drawn in Fig. 1, we must have
\begin{equation}
\lambda = {1 \over R}\,.
\end{equation}
The quantity $s_0$ is arbitrary and can be set equal to zero.  Then the path length $s$ will be measured from the intersection of the circle and a horizontal line through the circle's center.
\begin{figure}
\includegraphics[width=3.125truein]{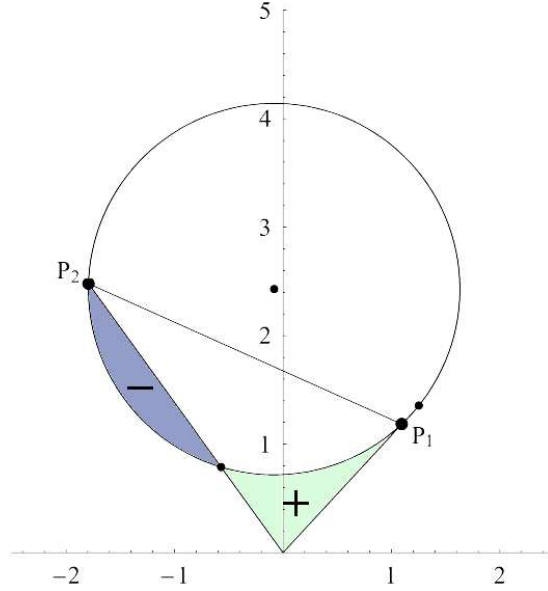}
\caption{Extremal path with zero net Sagnac discrepancy, for which there is a single region with negative contribution canceling the positive contribution. Here $P_1=(1.0950,1.1855)$, $P_2 = (-1.7930,2.4798)$, $\Psi=224.937\deg$, $R=1.71237$.  The shaded areas are $\pm 0.5045 $.}
\end{figure}
\subsubsection{Example with two contributing regions}
	In Fig. 2 we give another example of an extremal path on a rotating disc.  The circular path is described in a clockwise sense starting from the point $P_1$.  In this example, the straight line from the origin to $P_1$ does not cut off a section of the circular path.  Vanishing area results from the sum of the positive contribution, that builds up as the synchronization process passes from the initial point to the point $Q_2$, which is the intersection of the circle with a straight line from the origin to the final point $P_2$.  Positive contributions continue to build up until a point is reached where the velocity would point directly away from the origin--that is, where the tangent to the circle intersects the origin.  Negative contributions that build up thereafter, as the path approaches the final point, then cancel all the positive contributions.

\subsection{Polar Coordinates}

	To illustrate the similarities between extremal paths on a rotating disc and on a rotating sphere, we shall reformulate the problem on the disc using polar coordinates $(\rho,\phi)$, mentioning only the main points.  The path length is given by
\begin{equation}
\label{dssqrd2}
ds^2 = d\rho^2 + \rho^2 d\phi^2\,, 
\end{equation}
and the quantity to extremize is
\begin{equation}
W=\int_{P_1}^{P_2} L dt = \int_{P_1}^{P_2}\left(\sqrt{\dot \rho^2+\rho^2 \dot \phi^2} + {\lambda \over 2} \rho^2 \dot \phi \right) dt.
\end{equation}
The coordinate $\phi$ is cyclic, so the corresponding canonical variable $p_{\phi}$ is a constant of the motion:
\begin{equation}
p_{\phi} = \partial L / \partial \dot \phi= \rho^2 {d\phi \over ds} + {1 \over 2} \lambda \rho^2 = {\rm const.}=h\,,
\end{equation}
so 
\begin{equation}
\label{phidot}
{d\phi \over ds} = {h \over \rho^2} - {\lambda \over 2}.
\end{equation}
In order that the net projected area vanish, it follows from Eq. (\ref{phidot}) that $h$ and $\lambda$ cannot be of opposite sign.  In order that the net change in $\phi$ be positive, both these constants must be positive. 

The Euler-Lagrange equations have as a first integral the expression for the element of length, Eq. (\ref{dssqrd2}).  Eliminating $d\phi/ds$ from the length element gives
\begin{equation}
\left({d\rho \over ds} \right)^2 = 1-{h \over \rho^2} + h \lambda - {\lambda^2 \rho^2 \over 4}\,,
\end{equation}
and after some algebra, this can be rewritten
\begin{equation}
\label{drhosqrd}
\pm { d \rho^2 \over \sqrt{-{4 h^2 \over \lambda^2} + { 4(1+h\lambda) \rho^2 \over \lambda^2} - \rho^4}} = \lambda ds\,.
\end{equation}
The quantity inside the square root in the denominator of Eq. (\ref{drhosqrd}), considered as a function of the parameter $\rho^2$, is a parabola opening downward.  In a plot of this quantity as a function of $\rho^2$ the physical region is the first quadrant, limited by maximum and minimum values of $\rho^2$ where the denominator vanishes.  We shall denote these roots by $\rho_a^2$ and $\rho_b^2$, with:
\begin{eqnarray}
\label{rhoroots}
\rho_b^2 = {2 \over \lambda^2}\left( 1+ h \lambda + \sqrt{1 + 2 h \lambda}\right);\quad\quad
\rho_a^2 = {2 \over \lambda^2}\left( 1+ h \lambda - \sqrt{1 + 2 h \lambda}\right).
\end{eqnarray}
Both these roots are positive since $1+ h\lambda \ge \sqrt{1+2 h\lambda}$, which can be proved by squaring.  Thus the equation of motion for $\rho$ becomes
\begin{equation}
\pm { d \rho^2 \over \sqrt{(\rho_b^2-\rho^2)(\rho^2-\rho_a^2)}} = \lambda ds\,.
\end{equation}
The integration is standard and the solution of the equations of motion is
\begin{equation}
\label{rhosol1}
\rho^2 =  {2 \over \lambda^2}\left( 1+ h \lambda\right)+ {2 \over \lambda^2} \sqrt{1 + 2 h \lambda}\cos\left(\lambda(s-s'_0)\right),
\end{equation}
where $s'_0$ is an integration constant.  To identify the constants $h,\lambda$ in terms of the circular solutions discussed in the previous subsection, use  Eqs. (\ref{solutions1}) to calculate $\rho^2$: 
\begin{eqnarray}
\label{rhosol2}
x^2 + y^2 = \rho^2 = x_c^2+y_c^2 +R^2 +2 R(x_c \cos\left(\lambda(s-s_0)\right)-y_c \sin\left(\lambda(s-s_0)\right)\nonumber\\
=  x_c^2+y_c^2 +R^2 + 2 R \sqrt{x_c^2+y_c^2}\cos\left(\lambda(s-s'_0)\right). 
\end{eqnarray}
It follows that
\begin{equation}
\tan \left(\lambda s'_0\right) = { \tan \left(\lambda s_0\right) -y_c/x_c  \over 1+(y_c/x_c) \tan \left(\lambda s_0\right)}.
\end{equation}
Then identifying the coefficients in Eq. (\ref{rhosol2}) with those in Eq. (\ref{rhosol1}) gives
\begin{equation}
\lambda = {1 \over R};\quad\quad
h= {x_c^2+y_c^2 -R^2 \over 2 R}.
\end{equation}
The solutions given in Eqs. (\ref{solutions1}) and (\ref{rhosol1}) are equivalent.  The first-order differential equation, Eq. (\ref{drhosqrd}), is of the same form as the corresponding differential equation for extremal paths on a rotating sphere, which are discussed in the next section.
\section{Extremal Paths on a Rotating Sphere}
\subsection{Cylindrical Coordinates}

	On a rotating sphere, paths of extremal length for which the Sagnac effect vanishes can show new features which do not appear on the rotating disc.  In particular the presence of an upper and a lower hemisphere raises the question, what will a path look like which is required to cross the equator?  In other words, what is the shape of an isosynchronous path from say, an initial point below the equator, to a final point that is above the equator?  One might expect that if the specified endpoints are on the same hemisphere the shape of the path would bear some resemblance--when projected onto the equatorial plane--to an extremal path on the rotating disc.  However when the path must cross the equator some entirely new features arise.  In this paper we consider only paths which touch or cross the equator.

	We use cylindrical coordinates ($\rho,\phi,z$) to describe points on the sphere.  The sphere's radius is taken to be unity.  There is then a constraint
\begin{equation}
\label{sconstraint}
\rho^2+z^2=1,
\end{equation}
and we shall eliminate $\rho$ in favor of $z$ in constructing the Lagrangian.  We look for solutions in which $z$ passes through zero, corresponding to crossing the equator. 

	Differentiating Eq. (\ref{sconstraint}), it is straightforward to show that
\begin{equation}
\dot z^2 + \dot \rho^2 = {\dot z^2 \over 1 - z^2}\,.
\end{equation}
The element of path length on the sphere is then given by
\begin{equation}
\label{pathlength}
ds^2 =  {dz^2 \over 1 - z^2}+(1-z^2) d\phi^2\,,
\end{equation}
so the effective Lagrangian is
\begin{equation}
\label{lagrangian2}
L = \sqrt{{\dot z^2 \over 1-z^2}+(1-z^2)\dot \phi^2}+{1 \over 2}(1-z^2) \dot \phi\,.\end{equation}
Again the angular variable $\phi$ is cyclic, so the momentum canonically conjugate to $\phi$ is a constant of the motion which we denote by $h$.  Then this gives
\begin{equation}
\label{firstint}
{d\phi \over ds} = {h \over 1-z^2} - {\lambda \over 2}\,.
\end{equation}
This is of the same form as for the disc (see Eq. (\ref{phidot})).  Just as on a rotating disc, it is clear from Eq. (\ref{firstint}) that $\lambda$ and $h$ cannot have opposite signs for then $d\phi/ds$ could not change sign and the projected area could not vanish.  In order for the net change in $\phi$ to be positive, $h$ must be positive.  Therefore we shall assume that $\lambda$ and $h$ are both positive in our discussion. 

	One may proceed to derive the Euler-Lagrange equation for $z$ and integrate it, but it is more direct to use the constraint, Eq. (\ref{pathlength}), which is already a first integral of the equations of motion:
\begin{equation}
1 = (1-z^2)^{-1}\left({dz \over ds}\right)^2+(1-z^2)\left({ d \phi \over ds}\right)^2\,,
\end{equation}
and then using Eq. (\ref{firstint}), the equation is separable:
\begin{equation}
\label{secondint}
{dz^2 \over (1+h\lambda)(1-z^2) - h^2 - {\lambda^2 \over 4} (1-z^2)^2} = ds^2\,.
\end{equation}
The denominator of Eq. (\ref{secondint}) is a quadratic polynomial in the variable $(1-z^2)$.  In the allowed physical region this polynomial must be positive since the other factors in the equation are positive.  Consider then the roots of the polynomial\begin{equation}
F(x)= (1+h\lambda)(1-x) - h^2 - {\lambda^2 \over 4} (1-x)^2\,. 
\end{equation}	
These are at the values of x given by
\begin{equation} x_{\pm} = 1-{2 \over \lambda^2}(1+h\lambda)\pm {2 \over \lambda^2}\sqrt{1 + 2 h \lambda}\,.
\end{equation}
The smallest root has to be negative, otherwise $z=0$ cannot lie on the path.  Therefore we define 
\begin{eqnarray}
\label{zs}
z_b^2 = 1-{2 \over \lambda^2}(1+h\lambda)+ {2 \over \lambda^2}\sqrt{1 + 2 h \lambda}\ge 0;\\
-z_a^2 = 1-{2 \over \lambda^2}(1+h\lambda)- {2 \over \lambda^2}\sqrt{1 + 2 h \lambda}\le 0\,.
\end{eqnarray}
From the latter inequality it follows that the parameters $h$ and $\lambda$, in addition to both being positive,  are limited by 
\begin{equation}
(h-\lambda/2)^2 \le 1.
\end{equation}
Then Eq. (\ref{secondint}), after taking a square root, can be written
\begin{equation}
\label{finaleq}
\pm { dz \over \sqrt{(z_b^2-z^2)(z^2+z_a^2)}} = {\lambda \over 2} ds.
\end{equation}
For paths which don't cross the equator, the lesser root $x_-$ is not necessarily positive; we shall not discuss this case here.
\subsection{Solution in Terms of Elliptic Functions}

	The equation for the path, Eq. (\ref{finaleq}), can be written in terms of Jacobian elliptic functions.\cite{abramstegun}   Relevant properties of these functions are briefly summarized in the Appendix.  With these, it is straightforward to verify that a solution of Eq. (\ref{finaleq}) with the property that coordinate $z$ can pass through zero is 
\begin{equation}
\label{solnforz}
z(s) = {z_a z_b \over\sqrt{ z_a^2+z_b^2}} {\sn(u|m) \over \dn(u|m)}\,.
\end{equation}
where 
\begin{equation}
m = {z_b^2 \over z_a^2+z_b^2},\quad u = {\lambda \over 2}\sqrt{z_a^2+z_b^2}\,(s-s_0)\,.
\end{equation}
The constants $h$, $\lambda$ (or equivalently $z_a$, $z_b$) and the initial and final values of $u$ must be determined by satisfying the boundary conditions, and the condition that the projected area vanish.  
\subsubsection{Vanishing Sagnac Effect}
	The condition that the net Sagnac Effect vanish is most easily expressed by integrating Eq. (\ref{firstint}) between the endpoints.  Assuming that $u_1$ and $u_2$ are known, and since $du$ is proportional to $ds$, we have
\begin{equation}
{1 \over 2}\int_{u_1}^{u_2}(1-z^2)d\phi={1 \over 2} \left[{ds \over du}\right]
	\int_{u_1}^{u_2} du \left( h - {\lambda \over 2} +{\lambda \over 2}{z_a^2z_b^2 \over z_a^2 + z_b^2}{\sn^2(u|m) \over \dn^2(u|m)} \right)
\end{equation}
	The integral can be evaluated in terms of the standard elliptic integral of the second kind, $E(u|m)$, (see Appendix). 

	Then we find
\begin{equation}
0 = \bigg[\left(h-{\lambda \over 2}(1-z_a^2)\right)u +{\lambda \over 2}(z_a^2+z_b^2)\left(E(u|m)-{\sn(u|m)\cn(u|m) \over \dn(u|m)}\right)\bigg]_{u_1}^{u_2}\,.
\end{equation}
Also, for prescribed endpoints on the sphere, there will be a definite net increment in $\phi$ which from Eq. (\ref{firstint}) can be expressed by
\begin{equation}
\Delta \phi = \int_{u_1}^{u_2} d\phi = \left[{ds \over du}\right]\int_{u_1}^{u_2} du \left( {h \over 1-z^2 } -{\lambda \over 2}\right)\,.
\end{equation}
Using the solution, Eq. (\ref{solnforz}), this can be put in the form
\begin{equation}
\Delta \phi = C_1(u_2-u_1) + C_2\left(\Pi(n;u_2|m)-\Pi(n;u_1|m)\right)\,,
\end{equation}
where
\begin{eqnarray}
C_1&=&{1 \over \sqrt{z_a^2+z_b^2}}\left({2 h/\lambda \over 1+z_a^2} -1 \right)\,,
\\
C_2&=&{1 \over 1+z_a^2}{2 h z_a^2 /\lambda \over \sqrt{z_a^2+z_b^2}}\,,
\\
n&=&m(1+z_a^2)\,,
\end{eqnarray}
and where $\Pi(n;u|m)$ is the elliptic integral of the third kind (see Appendix).  We may show that the parameter $n$ is less than unity in the present case, by recalling that $1+h\lambda > \sqrt{1+2 h \lambda}$.  Then from the definition of $z_b$, Eq. (\ref{zs}), $z_b^2 <1$.  Thus $z_b^2 z_a^2 <z_a^2$, and
\begin{equation}
n={z_b^2(1+z_a^2)\over (z_a^2+z_b^2)}={z_b^2+z_b^2 z_a^2)\over (z_b^2+z_a^2)}<1.
\end{equation}
This makes the numerical calculation of $\Pi(n;u|m)$ somewhat easier.
\subsubsection{Example: Endpoints on the Equator}

	A special case arises when the two endpoints are on the equator, separated by azimuthal angle $\Delta \phi$.  Then obviously $u_1=0$ and $u_2=2K[m]$, because $\sn(u|m)$ must pass through a maximum and return to zero, that is, that the argument $u$ must increase by two quarter-periods.  Then
for a path which rises to positive $z$ and then returns to the equator we have 
\begin{equation}
u_2=2K[m]=2K\,.
\end{equation}
The condition that the Sagnac effect vanish becomes
\begin{equation}
0=\left(h - {\lambda \over 2}(1-z_a^2)\right) 2 K+{\lambda \over 2}(z_a^2+z_b^2)E(2K|m)\,.
\end{equation}
The net increment $\Delta \phi$ in $\phi$ will be
\begin{equation}
\Delta \phi = C_1 2 K + C_2 \Pi(n;2K|m)\,.
\end{equation}
These two highly transcendental equations must be solved simultaneously for $h$ and $\lambda$.  One solution is pictured in Figures 3 and 4 for the case $\Delta \phi = \pi/6$.  Figure 4 shows the path projected onto the equatorial plane.  The projection is not a conic section.
	
\begin{figure}
\includegraphics[width=3.125truein]{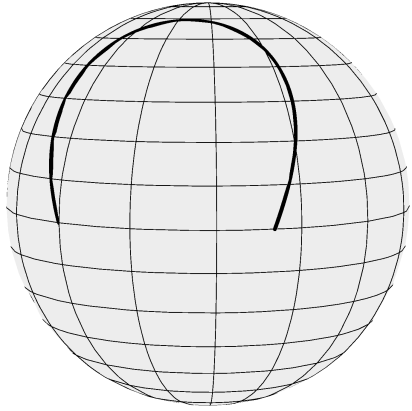}
\caption{Solution for endpoints on the equator separated by $\pi/6$.  The values of the parameters are $h=1.1284, \lambda = 3.1942$.} 
\end{figure}
\begin{figure}[ht]
\includegraphics[width=3.125truein]{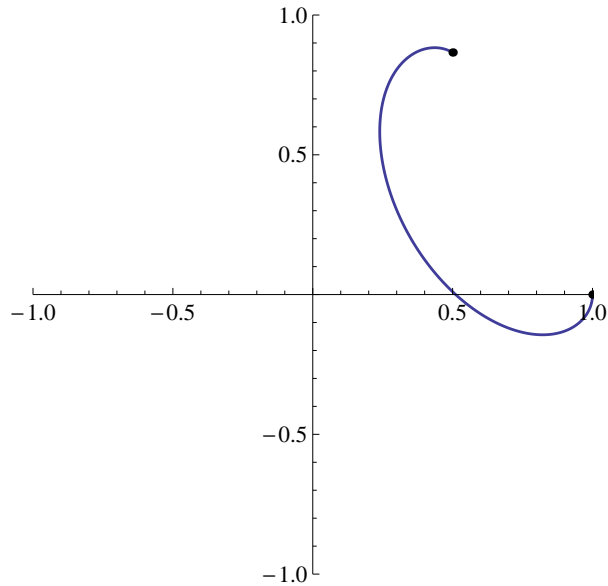}
\caption{Projection of the solution of  Fig. 3, onto the equatorial plane.}
\end{figure}

\subsubsection{Example with Loops}

	It is well-known that between two fixed endpoints on a torus there are many extremal paths.  Likewise there are many extremal paths on a sphere between two given endpoints, which satisfy the condition that the projected area vanish.  This possibility arises because the Jacobian elliptic functions such as $\sn(u|m)$ are periodic functions of the argument $u$.  For example, for the endpoints considered in the example illustrated in Figures 3 and 4, one could seek an extremal solution which rises to some positive value of $z$, crosses the equator and sinks to some negative value of $z$, and then returns to the equator.  This would correspond to setting $u_2=4K[m]$.  

Figures 5 and 6 give an example of such a path which crosses the equator three times.  The initial point is below the equator at $P_1 = (0.8,0,-0.6 )$ and the final point is above the equator at $P_2=(-0.3 ,\sqrt{0.27} ,+0.8 )$, separated from $P_1$ by an azimuthal angle $2\pi/3$.  The parameter values which solve the equations are $h=0.58227  $, $\lambda =1.7695 $.  When the path crosses the equator, $z=0$, and $d\phi/ds$ is negative.  Latitude vs. longitude of this path on the sphere's surface is plotted in Figure 6. 

A solution with the same endpoints, but which crosses the equator only once, can be constructed from parameter values $h=0.18395$, $\lambda=0.70195$.
\begin{figure}
\includegraphics[width=3.125truein]{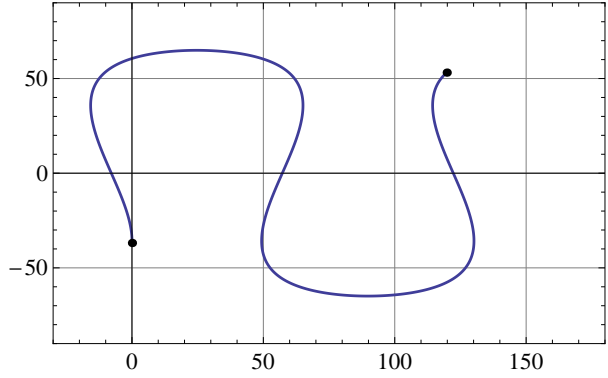}
\caption{Equatorial projection of extemal path from $(.8,0,-.6 )$ to $(-0.3 ,\sqrt{0.27} ,+0.8)$ that crosses the equator three times.}
\end{figure}
\begin{figure}
\includegraphics[width=3.125truein]{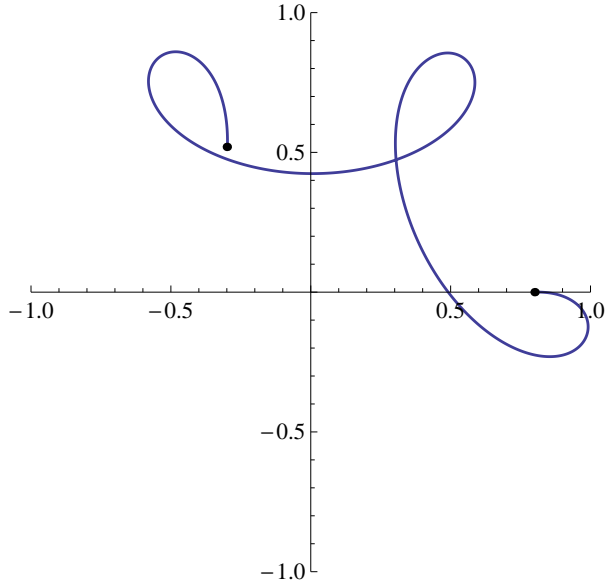}
\caption{Latitude-longitude plot of extremal path of Fig. 5.}
\end{figure}

\section{Conclusion}

	It has been shown here that between two given endpoints on the surface of a rotating sphere, there are an infinite number of paths for synchronization processes using either electromagnetic signals or slowly moving portable clocks, along which the Sagnac effect vanishes.  The problem can also be formulated purely as a mathematical problem: find paths of extremal length on a disc or sphere, for which the area swept out by a vector from the rotation axis to a point on the path, projected onto the equatorial plane, is zero.  The Sagnac effect has grown in practical importance in world-wide clock synchronization and navigation using the Global Positioning System.  While the issue addressed in this paper may not result in actual applications within modern synchronization or navigation systems, the problem is interesting and leads to some unexpected conclusions.  This problem can also be formulated and solved for a rotating ellipsoid, but the analytic solutions are so complicated as to be of little use.  

\begin{acknowledgments}
	This problem was suggested by a participant at the 1979 NBM-SAMSO Relativity Seminar in Boulder, Colorado.
\end{acknowledgments}	
\appendix*
\section{Properties of Elliptic Functions and Integrals}
	Here we list the most useful properties of the Jacobian elliptic functions, and the elliptic integrals of the first, second, and third kinds.  The notation is the same as that used in \cite{abramstegun}.  Consider the elliptic integral of the first kind:
\begin{equation}
F(u|m)=u=\int_0^{\phi} {d \theta \over (1-m \sin^2 \theta)^{1/2}}\,.
\end{equation}
Given $u$, one can solve for $\phi$, which is then called the amplitude:
\begin{equation}
\phi = \am(u|m)
\end{equation}
and the basic Jacobian elliptic functions are defined by:
\begin{eqnarray}
\sn(u|m)&&=\sin(\phi)\,,\\
\cn(u|m)&&=\cos(\phi)\,,\\
\dn(u|m)&&=\left(1-m \sin^2(\phi)\right)^{1/2}\,.
\end{eqnarray}
The dependence upon the parameter $m$ is frequently left implicit, as in writing $\sn\, u = \sin\, \phi,$ and so forth.

	The elliptic functions $\sn(u|m), \cn(u|m)$, and $\dn(u|m)$ satisfy the identities
\begin{eqnarray}
\label{ellipticdefs}
\sn(u|m)^2+ \cn(u|m)^2 = 1\,,\\
\dn(u|m)^2+m\, \sn(u|m)^2=1\,,
\end{eqnarray}
and their derivatives with respect to the argument $u$ are:
\begin{eqnarray}
\label{diffs}
{d \over du} \sn(u|m) = \cn(u|m) \dn(u|m)\\
{d \over du} \cn(u|m) = -\sn(u|m) \dn(u|m)\\
{d \over du} \dn(u|m) = -m\, \sn(u|m) \cn(u|m)\,.
\end{eqnarray}
 
	The elliptic integral of the second kind is
\begin{equation}
E(u|m) = \int_0^u \dn^2 w\,dw = \int_0^{sn(u|m)}{(1-m t^2)^{1/2} \over(1-t^2)^{1/2} }dt.
\end{equation}
The elliptic functions of the first and second kinds are said to be {\it complete} when the amplitude is $\pi/2$.  For example, the complete elliptic integral of the first kind is
\begin{equation}
K=K[m]=F(\pi/2|m)=\int_0^{\pi/2}(1-m\sin^2 \theta)^{-1} d\theta.
\end{equation}
Thus K is the real ``quarter-period" of the Jacobian elliptic function $\sn(u|m)$, such that $\sn(K|m)=1$.
The elliptic integral of the third kind depends on an additional parameter $n$:
\begin{equation}
\Pi(n;u|m) = \int_0^u (1-n\, \sn^2(w|m))^{-1} dw\,.
\end{equation} 
\bibliography{Ashby.bib}
\end{document}